\documentclass[conference]{IEEEtran}
\IEEEoverridecommandlockouts
\usepackage{cite}
\usepackage{amsmath,amssymb,amsfonts}
\usepackage{algorithmic}
\usepackage{graphicx}
\usepackage{textcomp}
\usepackage{xcolor}
\def\BibTeX{{\rm B\kern-.05em{\sc i\kern-.025em b}\kern-.08em
    T\kern-.1667em\lower.7ex\hbox{E}\kern-.125emX}}
\usepackage{threeparttable}
\usepackage{picinpar}
\usepackage{stfloats}
\usepackage{setspace}
\usepackage[all]{xy}
\usepackage{float}
\usepackage{amsmath}
\usepackage{array}
\usepackage{mdwmath}
\usepackage{mdwtab}
\usepackage{eqparbox}
\usepackage{algorithmic}
\usepackage{algorithm}
\usepackage{exscale}
\usepackage{relsize}
\usepackage{supertabular}
\usepackage{subfigure}
\usepackage{threeparttable}
\usepackage{caption}
\usepackage{booktabs}
\usepackage{multirow}
\usepackage{etoolbox}
\usepackage{makecell}
\usepackage{amssymb}
\usepackage{mathrsfs}
\usepackage{color}
\definecolor{purple}{RGB}{160,32,240}
\definecolor{darkred}{RGB}{255,0,255}
\usepackage{bm}
\usepackage{flushend}
\usepackage[justification=centering]{caption}

\renewcommand{\raggedright}{\leftskip=0pt \rightskip=0pt plus 0cm}

\begin{document}

\newtheorem{lemma}{Lemma}
\newtheorem{corol}{Corollary}
\newtheorem{theorem}{Theorem}
\newtheorem{proposition}{Proposition}
\newtheorem{definition}{Definition}
\newcommand{\e}{\begin{equation}}
\newcommand{\ee}{\end{equation}}
\newcommand{\eqn}{\begin{eqnarray}}
\newcommand{\eeqn}{\end{eqnarray}}

\title{Unsourced Massive Access-Based Digital\\
\vspace{-1.5mm}
Over-the-Air Computation for Efficient Federated\\
\vspace{-1.5mm}
Edge Learning}

\author{
  \IEEEauthorblockN{Li Qiao\IEEEauthorrefmark{1}, Zhen Gao\IEEEauthorrefmark{1}\IEEEauthorrefmark{2}\IEEEauthorrefmark{3}, Zhongxiang Li\IEEEauthorrefmark{1}\IEEEauthorrefmark{2}, and Deniz Gündüz\IEEEauthorrefmark{4}}
  
  \IEEEauthorblockA{\IEEEauthorrefmark{1}MIIT Key Laboratory of Complex-field Intelligent Sensing, Beijing Institute of Technology, Beijing 100081, China}
  
\IEEEauthorblockA{\IEEEauthorrefmark{2}Yangtze Delta Region Academy of Beijing Institute of Technology (Jiaxing), Jiaxing 314019, China}

\IEEEauthorblockA{\IEEEauthorrefmark{3}Advanced Technology Research Institute, Beijing Institute of Technology, Jinan 250307, China}

\IEEEauthorblockA{\IEEEauthorrefmark{4}Department of Electrical and Electronic Engineering, Imperial College London, London SW7 2AZ, U.K.}

Email: \{qiaoli, gaozhen16, lizhongxiang\}@bit.edu.cn, d.gunduz@imperial.ac.uk
\thanks{This work  received  funding  from  the  CHIST-ERA  project  SONATA (CHIST-ERA-20-SICT-004) funded by EPSRC-EP/W035960/1. For the purpose of open access, the authors have applied a Creative Commons Attribution (CCBY)  license  to  any  Author  Accepted  Manuscript  version  arising  from  this submission. This work was supported in part by the Natural Science Foundation of China (NSFC) under Grant 62071044 and Grant U2001210; in part by the Shandong Province Natural Science Foundation under Grant ZR2022YQ62; in part by the Beijing Nova Program; in part by the China Scholarship Council.}
}

\maketitle

\begin{abstract}
Over-the-air computation (OAC) is a promising technique to achieve fast model aggregation across multiple devices in federated edge learning (FEEL). In addition to the analog schemes, one-bit digital aggregation (OBDA) scheme was proposed to adapt OAC to modern digital wireless systems. However, one-bit quantization in OBDA can result in a serious information loss and slower convergence of FEEL. To overcome this limitation, this paper proposes an unsourced massive access (UMA)-based generalized digital OAC (GD-OAC) scheme. Specifically, at the transmitter, all the devices share the same non-orthogonal UMA codebook for uplink transmission. The local model update of each device is quantized based on the same quantization codebook. Then, each device transmits a sequence selected from the UMA codebook based on the quantized elements of its model update. At the receiver, we propose an approximate message passing-based algorithm for efficient UMA detection and model aggregation. Simulation results show that the proposed GD-OAC scheme significantly accelerates the FEEL convergences compared with the state-of-the-art OBDA scheme while using the same uplink communication resources.
\end{abstract}
\addtolength{\topmargin}{0.15in}


\IEEEpeerreviewmaketitle
\vspace{-1.5mm}
\section{Introduction}

With the growing deployment of Internet-of-Things (IoT), an increasing amount of data will be acquired by massive number of IoT devices \cite{JSAC-Editor}. Traditionally, collected data is offloaded to the cloud or a data center for data-driven machine learning (ML) applications \cite{AI-mag}. Recently, due to privacy concerns and the growing computation abilities of edge IoT devices (i.e., smartphones), wireless networks are pushing the deployment of centralized ML algorithms towards distributed learning frameworks \cite{AI-mag}. In the emerging federated edge learning (FEEL) framework, multiple edge IoT devices are coordinated by a central server to train an ML model using local datasets and computing resources \cite{Deniz2}. FEEL requires solving the distributed training problem taking into account the limited shared wireless resources and interference among multiple devices. Moreover, the model aggregation process in FEEL involves repeated uplink (UL) transmission of high-dimensional local gradients (or model updates) by tens to hundreds of devices, which places a heavy burden on the multiple access networks \cite{G_Zhu}.

To solve this problem, analog over-the-air computation (OAC) was introduced for efficient model aggregation \cite{Deniz, Deniz2, G_Zhu, Amiri}. Since the target of model aggregation is to calculate the average of all the local model updates rather than decoding each of the transmitted messages, the basic idea of analog OAC is to create and leverage inter-user interferences over the multiple access channel (MAC). Specifically, each element of the local model updates is precoded by the inverse of the UL channel gain and then modulated on the amplitude of the transmit waveform. After simultaneous transmission, the receiver directly acquires the sum of the local model updates of multiple devices based on the superposed waveform over MAC.

To further reduce the communication overhead, by exploiting sparsification and error accumulation, the dimension of the local model updates are reduced before analog transmission in \cite{Deniz}. After analog OAC, the receiver reconstructs the average of local model updates by using the approximate message passing (AMP) algorithm. 
A higher FEEL accuracy is achieved with limited communication resources in \cite{Yuxuan} by exploiting time-correlated sparsification on the local model updates \cite{Ozfatura}. In addition, compressive sensing (CS) techniques are adopted for communication efficient FEEL in \cite{GangLi} and \cite{CD-FL}. Different from \cite{Deniz, Deniz2, G_Zhu, Amiri, Yuxuan, GangLi, CD-FL,Ozfatura}, where all the participating devices are synchronized in time, misaligned analog OAC is considered in \cite{Yulin}.

On the other hand, most of the existing wireless networks adopt digital communication protocols (e.g., 3GPP standards) as well as hardware \cite{Kaibin}, and they may not be capable of employing an arbitrary modulation scheme, which is essential for the analog OAC schemes mentioned above. The authors of \cite{Kaibin} proposed a one-bit digital aggregation (OBDA) scheme for FEEL, which combines the ideas of sign stochastic gradient descent (signSGD) and OAC. Specifically, one-bit gradient quantization and digital modulation are adopted at devices, and majority-vote based gradient-decoding via OAC is employed at the central server. However, one-bit gradient quantization is too aggressive, resulting in slow convergence. On the other hand, if the local gradients (or model updates) are quantized using more bits, the superimposed digitally transmitted symbols at the receiver are no longer equal to the sum of the quantized gradients (or model updates). This necessities a novel digital OAC-based scheme with generalized quantization levels.

\begin{figure}[t]
 \vspace{-5mm}
     \centering
     \includegraphics[width = 0.77\columnwidth,keepaspectratio]
     {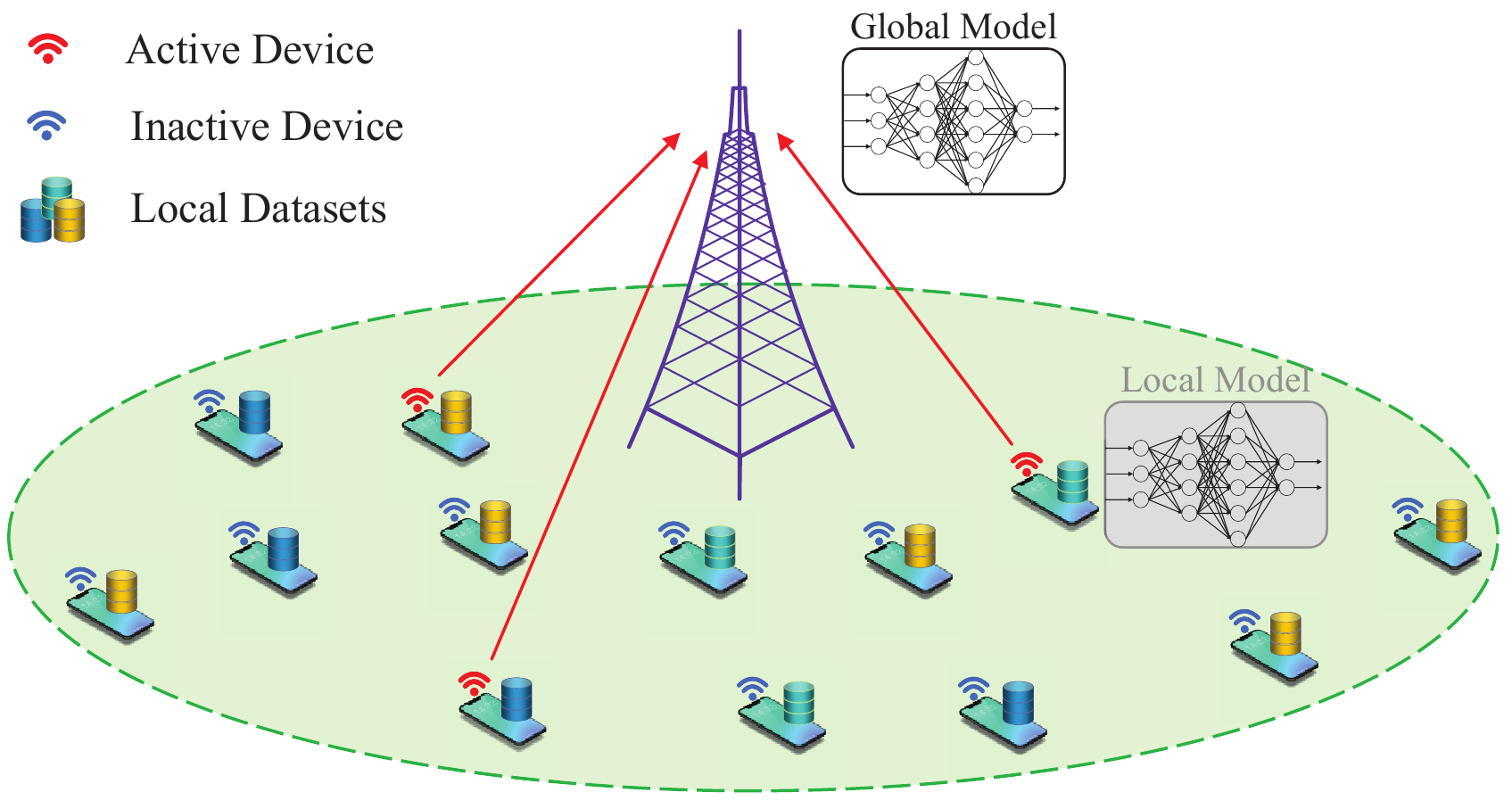}
     \captionsetup{font={footnotesize, color = {black}}, singlelinecheck = off, justification = raggedright,name={Fig.},labelsep=period}
     \caption{Illustration of the UMA-based FEEL scenario.}
     \label{fig1}
     \vspace{-8mm}
\end{figure}

Recently, grant-free random access has received attention to enable massive access of IoT devices with low signaling overhead and latency \cite{WuMag}, where devices can directly transmit their signals without the permission of the base station (BS). There are mainly two types of grant-free random access schemes, i.e., sourced massive access \cite{Malong, Qiao2, Mei, KeML_JSAC21, Qiao} and unsourced massive access (UMA) \cite{WuJSAC, Yury, Shao_UMA,MLK_UMA}. In the former, each device has its unique non-orthogonal preamble indicating its identity, and the BS performs active device detection and channel estimation, followed by data decoding \cite{Malong, Qiao2, Mei, KeML_JSAC21, Qiao}. As for UMA, all the devices adopt the same non-orthogonal UMA codebook. Transmitted bits are modulated by the index of the UMA codewords and decoded at the BS. The goal of the BS is to decode the list of messages, but not the identities of the active devices \cite{WuJSAC, Shao_UMA,Yury, MLK_UMA}. We remark that the identities of active devices are not necessary for model aggregation of FEEL either. Hence, our goal here is to redesign the modulation and decoding modules of UMA to tailor it for efficient digital model aggregation in FEEL across massive number of devices.

We propose an UMA-based generalized digital OAC (GD-OAC) scheme for communication efficient FEEL. Specifically, local model updates are quantized using the same quantization codebook. Then, the active devices select their transmit sequences (codewords) from the common UMA codebook based on their quantized model updates. Note that there is a one-to-one mapping between the quantization codebook and the non-orthogonal UMA codebook. Transmitted sequences overlap at the BS, which employs an AMP-based digital aggregation (AMP-DA) algorithm to calculate the average of the local model updates. Our simulation results verify that the proposed GD-OAC scheme is superior to the state-of-the-art OBDA scheme in terms of the test accuracy with the same UL communication resources.

\textit {Notation}: Boldface lower and upper-case symbols denote column vectors and matrices, respectively. For a matrix ${\bf A}$, ${\bf A}^T$, ${\left\| {\bf{A}} \right\|_F}$, $[{\bf{A}}]_{m,n}$ denote the transpose, Frobenius norm, the $m$-th row and $n$-th column element of ${\bf{A}}$, respectively. For a vector ${\bf x}$, $\| {\bf x} \|_p$ and $[{\bf x}]_{m}$ denote the ${l_p}$ norm and $m$-th element of ${\bf x}$, respectively. $|\Gamma|$ denotes the cardinality of the ordered set $\Gamma$. $\lfloor \cdot \rfloor$ rounds each element to the nearest integer smaller than or equal to that element. The marginal distribution $p\left([{\bf x}]_m\right)$ is denoted as $p\left([{\bf x}]_m\right)={\int}_{\backslash [{\bf x}]_m} p\left({\bf x}\right)$. $\mathcal{N}(x; \mu, \nu)$ (or $x\sim\mathcal{CN}(\mu, \nu)$) denotes the (complex) Gaussian distribution of random variable $x$ with mean $\mu$ and variance $\nu$. $[K]$ denotes the set $\{1,2,...,K\}$.

\vspace{-2mm}
\section{System Model}
As shown in Fig. \ref{fig1}, we consider $K$ edge devices served by a single antenna BS in a cellular system. Each device $k$, $k\in[K]$, has its own local dataset $\mathcal{D}_k$, which consists of labeled data samples. A common neural network model, represented by a parameter vector ${\bf w}\in\mathbb{R}^W$, is to be trained under the FEEL framework, coordinated by the BS. 

In each communication round of FEEL, e.g. the $t$-th round, the BS broadcasts the current global model ${\bf w}^t$ to the devices. The $k$-th device performs $E$ iterations of local stochastic gradient
descent (SGD) on ${\bf w}^t$ using its local dataset $\mathcal{D}_k$ \cite{Yuxuan}, which can be expressed as 
\begin{align}\label{LossGra}
    {\bf w}_{k, e}^t\!=\!{\bf w}_{k, e-1}^t\!-\!\eta\cdot\dfrac{1}{|\mathcal{D}_k|}\!\sum_{{\bm \varepsilon}\in\mathcal{D}_k} \!\nabla f({\bf w}_{k, e-1}^t,{\bm \varepsilon}),~e\in[E],
\end{align}
where ${\bf w}_{k, 0}^t={\bf w}^t$, $\nabla$ and $\eta$ denote the gradient operator and the learning rate, respectively, $f({\bf w}_{k, e-1}^t,{\bm \varepsilon})$ denotes the sample loss of model ${\bf w}_{k, e-1}^t$ on the training sample ${\bm \varepsilon}$. For convenience, we assume uniform sizes for local datasets, i.e., $|\mathcal{D}_k| = D$, $\forall k$. According to (\ref{LossGra}), the local model update can be obtained as ${\bf g}_k^t = {\bf w}_{k, E}^t - {\bf w}^t$, $\forall k,t$.

It is commonly adopted that only $K_a$ ($K_a\ll K$) active devices send their model updates to the BS in each round. If the local model updates can be perfectly obtained at the BS, the global model update ${\bf g}^t\in\mathbb{R}^W$ can be calculated as
\begin{align}\label{GlobalGra}
    {\bf g}^t=\dfrac{1}{K_a}\sum\nolimits_{k=1}^{K_a} {\bf g}_k^t,
\end{align}
where $K_a$ is usually unknown to the BS, due to the grant-free random access feature of IoT devices \cite{JSAC-Editor}. Then, the global model can be updated as
\begin{align}\label{GlobalWe}
    {\bf w}^{t+1}={\bf w}^{t} + {\bf g}^t.
\end{align}
Finally, the updated parameter vector ${\bf w}^{t+1}\in\mathbb{R}^W$ is sent back to the devices. The steps (\ref{LossGra}), (\ref{GlobalGra}), and (\ref{GlobalWe}) are iterated until a convergence condition is met.

It is clear from (\ref{GlobalGra}) that only the sum of the local model updates, rather than the individual values, is needed at the BS. The BS does not need to identify active devices either. These motivate the proposed UMA-based aggregation scheme presented in Section III.

\vspace{-3mm}
\section{Proposed UMA-Based GD-OAC Scheme}

\begin{figure}[t] 
\vspace{-4mm} 
\centering  
\includegraphics[width = 0.8\columnwidth]{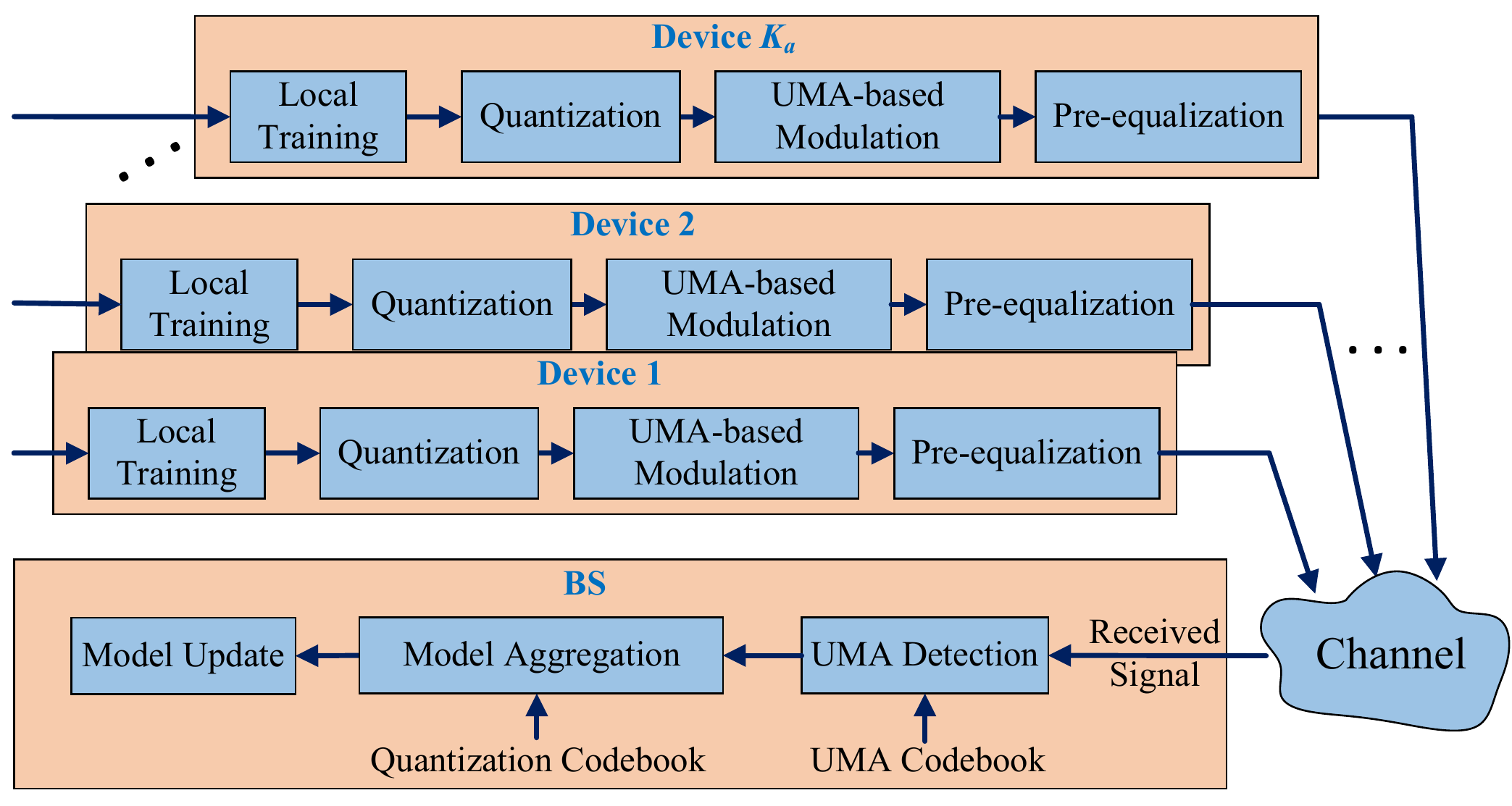}  
\captionsetup{font={footnotesize, color = {black}}, singlelinecheck = off, justification = raggedright,name={Fig.},labelsep=period}
\caption{The schematic diagram of the proposed UMA-based GD-OAC scheme.} 
\vspace{-8mm}
\label{fig2}
\end{figure}

Due to the large dimension of the parameter vector and the large number of devices that can potentially participate in the training, we have a communication bottleneck in FEEL. In this section, we propose to exploit UMA to reduce the communication overhead in FEEL. Next, we describe the transceiver design for the proposed UMA-based GD-OAC scheme in detail, illustrated in Fig. \ref{fig2}. For simplicity, we omit the index $t$ and focus on any one of the communication rounds.

\vspace{-2mm}
\subsection{Transmitter Design}
\vspace{-2mm}
\subsubsection{Quantization Design}
We consider the same quantization codebook for all of the devices, denoted as ${\bf U}=[{\bf u}_1, {\bf u}_2,...,{\bf u}_N]\in\mathbb{R}^{Q\times N}$, where $N=2^J$ denotes $J$-bit quantization with $N$ quantization codewords, $Q\geq 1$ ($Q\in\mathbb{N}$) denotes the length of each quantization codeword ${\bf u}_n\in\mathbb{R}^{Q}$, $\forall n\in[N]$. Note that $Q=1$ and $Q>1$ indicate the scalar and vector quantization (VQ), respectively.
\addtolength{\topmargin}{0.15in}

As for the $k$-th device, $\forall k\in[K_a]$, the indices of its quantized model update can be expressed as
\vspace{-1.5mm} 
\begin{align}\label{Quant}
    \overline{\bf g}_k=h\left({\bf g}_k, {\bf U}\right),
    \vspace{-2mm} 
\end{align}
where $\overline{\bf g}_k\in\mathbb{N}^{\overline{W}}$, $\overline{W}=W/Q$, and $h\left(\cdot, {\bf U}\right)$ is a function that maps ${\bf g}_k$ to $\overline{\bf g}_k$ based on the quantization codebook ${\bf U}$. For VQ ($Q>1$), supposing that $W$ can be divided by $Q$, $h\left(\cdot, {\bf U}\right)$ first reshapes ${\bf g}_k$ into a matrix with $Q$ rows and $\overline{W}$ columns, then each column is mapped to a quantization codeword with minimum Euclidean distance. Note that the dimension of each local model update can be reduced by using VQ. In addition, the $\overline{w}$-th element of $\overline{\bf g}_k$, denoted as $\overline{ g}_{k,\overline{w}}$, $\forall\overline{w}\in[\overline{W}]$, is an integer belonging to set $[N]$.

\subsubsection{UMA-based Modulation}
We consider the same non-orthogonal UMA codebook for all of the devices, denoted by ${\bf P} = [{\bf p}_1, {\bf p}_2,...,{\bf p}_N]\in\mathbb{R}^{L\times N}$, where $L$ is the length of each sequence ${\bf p}_n\in\mathbb{R}^{L}$, $\forall n\in[N]$. In the proposed scheme, there is a one-to-one mapping from ${\bf U}$ to ${\bf P}$. Hence, for the $k$-th device, the $\overline{w}$-th element of its quantized model update is modulated to the sequence ${\bf p}_{\overline{ g}_{k,\overline{w}}}\in\mathbb{R}^{L}$.

\subsubsection{UL Massive Access Model}
We consider a time division duplex system, where the UL and downlink (DL) channels are commonly considered the same due to channel reciprocity \cite{Mei}. Before each UL communication round of FEEL, the BS first broadcasts a pilot signal and each active device estimates its DL channel 
 $\widehat{h}_k^{\rm DL}$, $k\in[K_a]$, based on the pilot signal, then active devices start their UL transmissions with pre-equalization \cite{Mei}. For the $\overline{w}$-th element of $\overline{\bf g}_k$, $\forall k, \overline{w}$, the received signal ${\bf y}\in\mathbb{R}^{L}$ at the BS can be expressed as
 \vspace{-1.5mm} 
\begin{align}\label{Quant}
    {\bf y}_{\overline{w}}=\sum_{k=1}^{K_a}h_k^{\rm UL}\dfrac{1}{\widehat{h}_k^{\rm DL}}{\bf p}_{\overline{ g}_{k,\overline{w}}} + {\bf z}_{\overline{w}}={\bf P}{\bf x}_{\overline{w}} + {\bf z}_{\overline{w}},
    \vspace{-1.5mm} 
\end{align}
where $h_k^{\rm UL}\sim\mathcal{CN}(0,1)$ denotes the UL channel gain of the $k$-th active device, each element of the noise vector ${\bf z}_{\overline{w}}\in\mathbb{R}^{L}$ obeys the independent and identically distributed (i.i.d.) Gaussian distribution with zero mean and variance $\sigma^2$. Here, we will assume perfect estimation of the DL channel for simplicity, i.e., $\widehat{h}_k^{\rm DL} = h_k^{\rm UL}$. Also, ${\bf x}_{\overline{w}}\in\mathbb{N}^{N}$, $\forall \overline{w}\in[\overline{W}]$, is the equivalent transmit signal vector satisfying the following properties:
\vspace{-1.5mm} 
\begin{align}\label{TransX}
    \|{\bf x}_{\overline{w}}\|_1=K_a,~\|{\bf x}_{\overline{w}}\|_0\leq K_a,~x_n^{\overline{w}}\in \Omega,~\forall n\in[N],
\end{align}
where ${\bf x}_{\overline{w}}=[x_1^{\overline{w}}, x_2^{\overline{w}},...,x_N^{\overline{w}}]^T$ and $\Omega=\{0, [K_a]\}$. According to this notation, the $n$-th sequence ${\bf p}_n$ is transmitted by $x_n^{\overline{w}}$ active devices, or equivalently the $n$-th quantization codeword appears at $x_n^{\overline{w}}$ devices, which indicates how to conduct the model aggregation.

\vspace{-3mm}
\subsection{Receiver Design}
\subsubsection{UMA Detection}
To achieve higher quantization accuracy, especially for VQ, the number of quantization bits $J$ can be set very large, e.g., more than 8 bits. In this case, $N=2^J$ can be far greater than $K_a$, which motivates us to employ CS-based algorithms for efficient UMA detection with reduced communication overhead $L$ \cite{WuMag}. Note that the UMA codebook is known at the BS. The details of the proposed CS-based detection algorithm will be illustrated in Section IV.
\addtolength{\topmargin}{-0.05in}

\subsubsection{Model Aggregation}
After UMA detection, we can obtain the estimate of the equivalent transmit signals, denoted by $\widehat{\bf x}_{\overline{w}}\in\mathbb{N}^{N}$, $\forall \overline{w}$. According to (\ref{TransX}), the number of active devices $\widehat{K}_a$ can be estimated based on $\widehat{\bf x}_{\overline{w}}$, which will be detailed in Section IV-C. Then, according to (\ref{GlobalGra}), we can obtain the global model update corresponding to $\overline{g}_{k,\overline{w}}$ as
\begin{align}\label{GlobalEst}
    \widehat{\bf g}_{\overline{w}}=\dfrac{1}{\widehat{K}_a}{\bf U}\widehat{\bf x}_{\overline{w}},
\end{align}
where $\widehat{\bf g}_{\overline{w}}\in\mathbb{R}^{Q}$. Hence, the whole global model update can be obtained as $\widehat{\bf g}=[(\widehat{\bf g}_{\overline{1}})^T,(\widehat{\bf g}_{\overline{2}})^T,...,(\widehat{\bf g}_{\overline{W}})^T]^T\in\mathbb{R}^{W}$.

\subsubsection{Model Update}
According to (\ref{GlobalWe}) and (\ref{GlobalEst}), the BS can update the weight of the ML model as
\begin{align}\label{GlobalWeEs}
    \widehat{\bf w}^{t+1}=\widehat{\bf w}^{t} + \widehat{\bf g}^t.
\end{align}
Then, the updated model $\widehat{\bf w}^{t+1}$ is broadcast to all the devices.

\vspace{-2mm}
\section{Proposed CS-Based Detection and Model Aggregation Algorithm}
In this section, we will first introduce the CS-based problem formulation. Then, we propose the AMP-based algorithm for efficient UMA detection and $K_a$ estimation. Finally, we summarize the proposed AMP-DA algorithm.

\vspace{-2.5mm}
\subsection{Problem Formulation}
For simplicity, we omit the subscript $\overline{w}$ in this section and focus on any round of the UL transmission. To solve the UMA detection problem in (\ref{Quant}), we aim to minimize the mean square error between ${\bf y}$ and ${\bf Px}$. Equivalently, we can calculate the posterior mean of ${\bf x}$ under the Bayesian framework \cite{Qiao2}. The posterior mean of $x_n$, $\forall n\in [N]$, can be expressed as
\vspace{-2mm}
\begin{align}\label{eq:Pmean}
    \widehat{x}_n= {\int} x_n p\left(x_n| {\bf{y}}\right)dx_n,
\end{align}
where $p(x_n| {\bf{y}})$ is the marginal distribution of $p({\bf x}| {\bf{y}})$:
\begin{align}\label{eq:Marginal}
    p\left(x_n| {\bf{y}}\right)= \mathlarger{\int}_{\backslash x_n} p\left({\bf x}| {\bf{y}}\right).
\end{align}

As discussed in Section III-B, to achieve higher quantization accuracy, more quantization bits are needed for higher dimensional VQ. This can result in a large dimension of $N$, which demands an efficient algorithm to calculate the marginal distribution $p(x_n| {\bf{y}})$.

\subsection{Approximate Message Passing (AMP)}
To address this issue, the AMP algorithm can be adopted to obtain the approximate marginal distributions with relatively low complexity \cite{Meng}. According to the AMP algorithm, we can approximately decouple (\ref{Quant}) into $N$ scalar problems as
\begin{equation}\label{eq:Decoupling}
\begin{array}{l}
{\bf y}={\bf P}{\bf{x}} +{\bf{z}}\; \to \;r_n = x_n + z_n,
\end{array}
\end{equation}
where $n\in[N]$, $r_n$ is the mean of $x_n$ estimated by the AMP algorithm, and $z_n\sim {\cal N}(z_n;0,\varphi_n)$ is the associated noise with zero mean and variance $\varphi _n$ \cite{Malong}. Hence, according to Bayes's theorem, the posterior distribution of $x_n$ can be expressed as
\begin{align}\label{eq:Bayes}
    p\left(x_n| {\bf{y}}\right) \approx p\left(x_n| r_n\right)= \dfrac{1}{p\left(r_n\right)}p\left(r_n|x_n\right)p\left(x_n\right),
\end{align}
where $``\approx"$ is due to the AMP approximation, and
\begin{align}\label{eq:BayesDetail}
    p\left(r_n|x_n \right)&= {\cal N}(r_n;x_n,\varphi_n),\\
    p\left(r_n \right)&= \sum\nolimits_{x_n\in\Omega}p\left(r_n|x_n\right)p\left(x_n\right).
\end{align}

Furthermore, according to (\ref{TransX}), we can model the prior distribution of $x_n$ as 
\begin{align}\label{eq:prior}
    p\left(x_n \right)= \left(1-a_n \right)\delta(x_n) + \dfrac{a_n}{K_a}\sum\nolimits_{s\in[K_a]}\delta(x_n-s),
\end{align}
where the sparsity indicator $a_n = 0$ if $x_n=0$, otherwise $a_n = 1$. Also, in (\ref{eq:prior}), if $a_n = 1$, we assume that $x_n$ can be any element from set $[K_a]$ with equal probability. Note that this assumption is also an approximation since the BS cannot obtain the actual distribution of each element $x_n$, $\forall n\in[N]$.  

According to (\ref{eq:Bayes})$-$(\ref{eq:prior}), the posterior mean and variance of $x_n$, $\forall n\in[N]$, denoted by ${\widehat x}_n$ and ${\widehat v}_n$, respectively, can be evaluated as
\begin{align}
\label{eq:postmean}{\widehat x}_n&=\sum\nolimits_{x_n\in\Omega}x_n p(x_n| {\bf{y}})d x_n,\\
\label{eq:postvar}{\widehat v}_n&=\sum\nolimits_{x_n\in\Omega}|x_n|^2 p(x_n| {\bf{y}})d x_n - |{\widehat x}_n|^2.
\end{align}

In addition, in the $i$-th AMP iteration, $(r_n)^i$ and $(\varphi_n)^i$, $\forall n$, of (\ref{eq:Decoupling}) are updated as
\begin{align}
\label{eq:UpdateSigma}(\varphi_n)^i&=\left({\sum\nolimits_{l=1}^{L}\dfrac{\left|[{\bf p}_n]_{l}\right|^2}{\sigma^2+(V_l)^i}}\right)^{-1},\\
\label{eq:UpdateR} (r_{n})^{i}&=(\widehat{{x}}_{n})^{i}+(\varphi_{n})^{i}\sum\nolimits_{l=1}^{L}\dfrac{[{\bf p}_n]_{l}\left({\left[{\bf y}\right]_l-(Z_{l})^{i}}\right)}{\sigma^2+(V_{l})^{i}},
\end{align}
where $(V_{l})^{i}$ and $(Z_{l})^{i}$, $\forall l$, are updated as
\begin{align}
\label{eq:UpdateV} (V_{l})^{i}&=\sum\nolimits_{n=1}^N\left|[{\bf P}]_{l,n}\right|^2 (\widehat{v}_{n})^{i},\\
\label{eq:UpdateZ} (Z_{l})^{i}&=\sum\nolimits_{n=1}^N [{\bf P}]_{l,n}(\widehat{x}_{n})^{i}-(V_{l})^{i}\dfrac{\left[{\bf y}\right]_l-(Z_{l})^{i-1}}{\sigma^2+(V_{l})^{i-1}},
\end{align}
while $(\cdot)^i$ denotes its argument in the $i$-th AMP iteration. For further details of the AMP update rules in (\ref{eq:UpdateSigma})$-$(\ref{eq:UpdateZ}), we refer the readers to \cite{Meng}.

\subsection{Parameter Estimation}
\subsubsection{Estimation of $K_a$}
According to (\ref{eq:postmean}), we can obtain the estimated transmit signal vector $\widehat{\bf x}_{\overline{w}}=[\widehat{x}_{1}^{\overline{w}},\widehat{x}_{2}^{\overline{w}},...,\widehat{x}_{N}^{\overline{w}}]^T$, $\forall \overline{w}\in[\overline{W}]$. As indicated in (\ref{TransX}), $\|\widehat{\bf x}_{\overline{w}}\|_1$ should equal to $K_a$ under perfect UMA detection. To improve the robustness, we estimate $\widehat{K}_a$ as follows
\begin{align}\label{eq:KaEst}
    \widehat{K}_a \!=\! \Psi\left\{\left\lfloor\left\|\widehat{\bf x}_{\overline{w}}\right\|_1 \!\!+\! \frac{1}{2}\right\rfloor, \overline{w}\in[\overline{W}]\right\},
\end{align}
where function $\Psi\left\{\cdot\right\}$ calculates the most frequently occurring element of its argument. The intuition of (\ref{eq:KaEst}) is similar to majority voting to improve the detection accuracy of $K_a$.

\subsubsection{Estimation of the sparsity indicators}
The unknown sparsity indicators, denoted by $a_n$, $\forall n\in[N]$, can be obtained by using the expectation maximization (EM) algorithm. The EM algorithm update rules are as follows
\begin{align}\label{eq:EM} 
(a_n)^{i+1}={\rm arg}\max\limits_{a_n}\mathbb{E}\left\{{{\rm ln}~p\left({{\bf x, y}}\right)|{\bf y};(a_n)^i}\right\},
\end{align}
where $(a_n)^i$ denotes the sparsity indicator in the $i$-th iteration, $\mathbb{E}\{ \cdot|{\bf y};(a_n)^i\}$ represents the expectation conditioned on the received signal ${\bf y}$ under $(a_n)^i$. Hence, according to (\ref{eq:Bayes}), the sparsity indicators can be obtained as
\begin{align}\label{eq:EMACT}
(a_n)^{i+1}=\sum\nolimits_{x_n\in[K_a]}p\left( x_n|{\bf y};(a_n)^i\right).
\end{align}
The derivations of the EM update rule can be found in \cite{Malong}.

\addtolength{\topmargin}{-0.04in}
\subsection{Proposed AMP-DA Algorithm}

Based on (\ref{eq:postmean})-(\ref{eq:KaEst}) and (\ref{eq:EMACT}), we summarize the proposed AMP-DA algorithm in {\bf Algorithm} 1. The details are explained as follows.

In line \ref{A1:initial}, we initialize the sparsity indicators $a^{\overline w}_n$, the variables $V^{\overline w}_l$, $Z^{\overline w}_l$, the posterior mean ${\widehat x}_{n}^{\overline w}$, and posterior variance ${\widehat v}_{n}^{\overline w}$, $\forall \overline{w}, n, l$. The iteration starts in line \ref{A1:T0}. Specifically, lines \ref{A1:AMP-S}-\ref{A1:denoising} correspond to the AMP operation. In the $i$-th iteration of the AMP decoupling step (line \ref{A1:decoupling}), $V^{\overline w}_l$, $Z^{\overline w}_l$, $\varphi_{n}^{\overline w}$, and $r_{n}^{\overline w}$, $\forall \overline{w}, n, l$, are calculated according to (\ref{eq:UpdateV}), (\ref{eq:UpdateZ}), (\ref{eq:UpdateSigma}), and (\ref{eq:UpdateR}), respectively. Furthermore, a damping parameter $\tau$ is adopted in line \ref{A1:damping} to prevent the algorithm from diverging\cite{Malong}. In addition, in the AMP denoising step (line \ref{A1:denoising}), we calculate the posterior mean ${\widehat x}_{n}^{\overline w}$ and the corresponding posterior variance ${\widehat v}_{n}^{\overline w}$, $\forall \overline{w}, n$ of the $i$-th iteration, by using (\ref{eq:postmean}) and (\ref{eq:postvar}), respectively. Accordingly, EM operation is used to update the sparsity indicators $a^{\overline w}_n$ in line \ref{A1:EM-ACT}. Then, the iteration restarts in line \ref{A1:AMP-S} until the maximum iteration number $T_0$ is reached. Otherwise, if line \ref{A1:if} is triggered, i.e., the normalized mean square error between continuous $\widehat{\bf x}_{\overline{w}}$ is smaller than the predefined threshold $\epsilon$, the iteration ends in advance. After the iteration, we acquire the estimated equivalent transmit signal $\widehat{\bf x}_{\overline{w}}$, ${\forall \overline{w}}$, in line \ref{A1:Xest}. According to (\ref{eq:KaEst}), in line \ref{A1:KaEst}, we obtain the number of active devices $\widehat{K}_a$ based on $\widehat{\bf x}_{\overline{w}}$, ${\forall \overline{w}}$. Finally, according to (\ref{GlobalEst}), we obtain $\widehat{\bf g}_{\overline{w}}$ in line \ref{A1:DA} by using $\widehat{\bf x}_{\overline{w}}$ and $\widehat{K}_a$, ${\forall \overline{w}}$. Hence, the estimated global model update is acquired in line \ref{A1:DAofALL} as $\widehat{\bf g}=[(\widehat{\bf g}_{\overline{1}})^T,(\widehat{\bf g}_{\overline{2}})^T],...,(\widehat{\bf g}_{\overline{W}})^T]^T$, which realizes the generalized digital model aggregation process. Also, by recalling (\ref{GlobalWeEs}), the weight of the ML model will be updated using $\widehat{\bf g}$.

\begin{algorithm}[t]
\algsetup{linenosize=\footnotesize} \footnotesize
\color{black}
\caption{Proposed AMP-DA Algorithm}\label{Algorithm:1}
\begin{algorithmic}[1] 
\raggedright 
\REQUIRE The received signals ${\bf y}_{\overline{w}}\in \mathbb{R}^{L}$, $\forall \overline{w}\in[\overline{W}]$, the UMA codebook ${\bf P}\in\mathbb{R}^{L\times N}$, the quantization codebook ${\bf U}\in\mathbb{R}^{Q\times N}$, the noise variance ${\sigma^2}$, the maximum iteration number $T_0$, the damping parameter $\tau$, and the termination threshold $\epsilon$.
\ENSURE The estimated global model update $\widehat{\bf g}$.
\STATE ${\forall \overline{w}, n, l}$: We initialize the iterative index $i$~$=$~1, the sparsity indicators $(a^{\overline w}_n)^1\!=\!0.5$, $(Z^{\overline w}_l)^1\!=\!\left[{\bf y}_{\overline{w}}\right]_l$, $(V^{\overline w}_l)^1\!=\!1$, $({\widehat x}_{n}^{\overline w})^1\!=\!0$, and $({\widehat v}_{n}^{\overline w})^1\!=\!1$;
\label{A1:initial}
\FOR {$i=2$ to $ T_0$}
\label{A1:T0}
\STATE \textbf{\textbf{\%}AMP operation:}
\label{A1:AMP-S}
\STATE ${\forall \overline{w}, n, l}$: Compute $(V^{\overline w}_l)^i$, $(Z^{\overline w}_l)^i$, $(\varphi^{\overline w}_n)^i$, and $(r^{\overline w}_n)^i$ by using (\ref{eq:UpdateV}), (\ref{eq:UpdateZ}), (\ref{eq:UpdateSigma}), and (\ref{eq:UpdateR}), respectively;~~\{Decoupling step\}
\label{A1:decoupling}
\STATE ${\forall \overline w, l}$: $(V^{\overline w}_{l})^i=\tau(V^{\overline w}_{l})^{i-1}+(1-\tau)(V^{\overline w}_{l})^i$, \\
$(Z^{\overline w}_{l})^t=\tau(Z^{\overline w}_{l})^{i-1}+(1-\tau)(Z^{\overline w}_{l})^i$;~~\{Damping\}
\label{A1:damping}
\STATE ${\forall \overline{w}, n}$: Compute $({\widehat x}_{n}^{\overline{w}})^i$ and $({\widehat v}_{n}^{\overline{w}})^i$ by using (\ref{eq:postmean}) and (\ref{eq:postvar}), respectively;~~\{Denoising step\}
\label{A1:denoising}
\STATE \textbf{\textbf{\%}Parameter update:}
\label{A1:EM-S}
\STATE ${\forall \overline{w}, n}$: Update the sparsity indicators $(a^{\overline{w}}_n)^i$ by using (\ref{eq:EMACT});
\label{A1:EM-ACT}
\STATE $i = i+ 1;$
\IF{$\left[\sum\limits_{\overline{w}}\left\| (\widehat{\bf x}_{\overline{w}})^i-(\widehat{\bf x}_{\overline{w}})^{i-1} \right\|_F/\left\| (\widehat{\bf x}_{\overline{w}})^{i-1} \right\|_F\right]/\overline{W}<\epsilon$}
\label{A1:if}
\STATE ${\bf break}$;~~\{End the iteration\}
\label{A1:break}
\ENDIF
\label{A1:endif}
\ENDFOR
\label{A1:endfor}
\STATE ${\forall \overline{w}}$:The estimated equivalent transmit signal $\widehat{\bf x}_{\overline{w}}=(\widehat{\bf x}_{\overline{w}})^i$;
\label{A1:Xest}
\STATE \textbf{\textbf{\%}Estimate the number of active devices:}
\STATE According to (\ref{eq:KaEst}), we obtain $\widehat{K}_a$;
\label{A1:KaEst}
\STATE \textbf{\textbf{\%}Model aggregation:}
\STATE ${\forall \overline{w}}$:According to (\ref{GlobalEst}), we obtain $\widehat{\bf g}_{\overline{w}}$ by using $\widehat{\bf x}_{\overline{w}}$ and $\widehat{K}_a$;
\label{A1:DA}
\STATE The estimated global model update $\widehat{\bf g}=[(\widehat{\bf g}_{\overline{1}})^T,(\widehat{\bf g}_{\overline{2}})^T],...,(\widehat{\bf g}_{\overline{W}})^T]^T$.
\label{A1:DAofALL}
\end{algorithmic}
\end{algorithm}

\section{Simulation Results}

In this section, we evaluate the performance of the proposed UMA-based GD-OAC scheme by considering FEEL for the image classification task on the commonly adopted CIFAR-10 dataset \cite{Sun2}. CIFAR-10 dataset has a training set of 50,000 and a test set of 10,000 colour images of size 32 × 32 belonging to 10 classes. Consider $K=100$ devices in the cell collaboratively training a convolutional neural network (same structure as in \cite{Sun2}) with 258898 (i.e., $W \approx 2.6\times10^5$) parameters. We consider non-i.i.d. local training data across the devices as follows \cite{Yulin}: 1) Each one of the 100 devices is randomly assigned 300 samples from the dataset; 2) The remaining 20,000 samples are sorted by their labels and grouped into 100 shards of size 200. Then, each device is assigned one shard. Furthermore, the local training iterations $E=10$, the mini-batch size is 512, and the learning rate is $\eta=0.001$. The metric to evaluate the performance is the {\it test accuracy} of the model on the test dataset.

As for the wireless transmission, we assume around 10\% of the devices are active in each round, where $K_a$ is uniformly generated from 7 to 13. Since the receiver does not exactly have $K_a$, $\Omega=\{0, [0.2\times K]\}$, shown in (\ref{TransX}), is considered as a prior information. The signal-to-noise ratio is set to 20 dB. For the proposed AMP-DA algorithm, the damping parameter is set to $\tau=0.3$ and the termination threshold is $\epsilon=1\times10^{-5}$. The elements of UMA codebook ${\bf P}$ obey i.i.d. Gaussian distribution. In each communication round, K-means algorithm \cite{PRML} is used to generate the quantization codebook ${\bf U}$ based on the local model update of one of the active devices. 
 
We consider two benchmark schemes as follows. 1) {\bf Perfect aggregation (PA)}: The transmitter is the same as the proposed GD-OAC scheme, while we consider perfect model aggregation at the receiver. 2)  {\bf OBDA}: The state-of-the-art digital OAC scheme with one-bit quantization \cite{Kaibin}. Note that the OBDA scheme needs $W$ UL communication resources in each round. As for the proposed GD-OAC scheme, the communication resources required in each round is $L\times\frac{W}{Q}=\frac{L}{Q}\times W$. To realize the same UL communication resources, we set $L=Q$.

\begin{figure}[t]  
\vspace{-9mm}
\centering  
\includegraphics[width = 0.95\columnwidth]{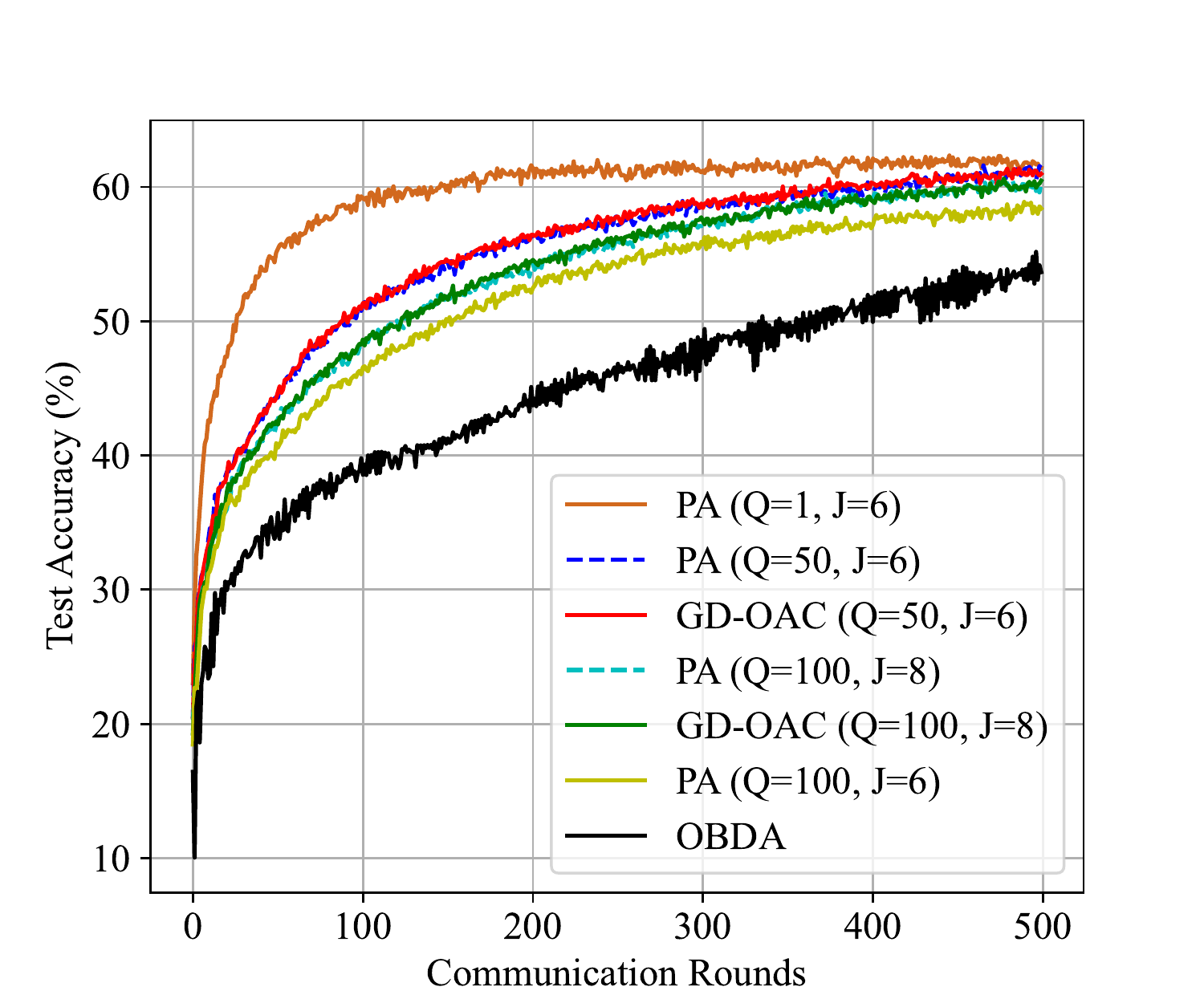}  
\captionsetup{font={footnotesize, color = {black}}, singlelinecheck = off, justification = raggedright,name={Fig.},labelsep=period}
\caption{Performance comparison of the proposed GD-OAC scheme and the benchmark schemes for training a model for the CIFAR-10 classification task.}
\vspace{-8mm}
\label{fig3}
\end{figure}

As shown in Fig. \ref{fig3}, by using the same communication resources, the proposed GD-OAC scheme significantly outperforms the OBDA scheme in terms of test accuracy. The reason is that one-bit quantization results in a severe quantization loss. Also, it can be seen that the proposed GD-OAC scheme achieves nearly the same accuracy as that of the PA schemes despite using limited communication resources. It is also observed that a larger $Q$ results in a slower convergence with fixed $J$, while increasing the number of quantization bits $J$ can improve the speed of convergence. Hence, there is a trade-off between the compression ratio (i.e., $\frac{1}{Q}$) and the test accuracy.

\section{Conclusions}
This paper proposed a UMA-based GD-OAC scheme for communication efficient FEEL across a large number of edge  devices. The main idea of the proposed solution relies on the fact that, in FEEL, the BS does not need the identities of the transmitting devices for model aggregation. The local model updates are first quantized based on a common quantization codebook, then each quantized element is modulated into a transmit sequence selected from the common non-orthogonal UMA codebook. These transmit sequences from different devices overlap at the BS. Then, the proposed AMP-DA algorithm can efficiently calculate the average of the local model updates. Due to the compression induced by VQ and the flexibility of quantization levels, simulation results verified that the proposed GD-OAC scheme significantly outperforms the OBDA scheme in terms of convergence speed.


\end{document}